\documentclass[journal]{IEEEtran}
\pdfoutput=1
\usepackage{ifpdf}
\usepackage{subfigure}
\usepackage{graphicx}
\DeclareGraphicsExtensions{.pdf,.jpg,.png}
\usepackage[cmex10]{amsmath}
\usepackage{amssymb}
\usepackage{array}
\usepackage{mdwmath}
\usepackage{mdwtab}
\usepackage{pslatex}
\usepackage{xcolor}
\usepackage{textcomp}
\usepackage{placeins}
\usepackage{eqparbox}
\usepackage{url}
\usepackage{stfloats}
\usepackage{multicol}
\usepackage{subfigure}
\usepackage{float}
\usepackage{algorithmicx}
\usepackage{algpseudocode}
\usepackage{amsthm}
\usepackage{comment}
\usepackage{todo}
\usepackage{enumitem}
\usepackage{gensymb}

\begin{document}

\title{Future of Ultra-Dense Networks Beyond 5G: Harnessing Heterogeneous Moving Cells}

\author{Sergey~Andreev,~\IEEEmembership{Senior~Member,~IEEE,}
        Vitaly~Petrov,~\IEEEmembership{Student~Member,~IEEE,}\\
        Mischa~Dohler,~\IEEEmembership{Fellow,~IEEE,}
        and~Halim~Yanikomeroglu,~\IEEEmembership{Fellow,~IEEE}
\thanks{This work was supported by the Academy of Finland (Project PRISMA). The work of V. Petrov was supported in part by the Nokia Foundation and in part by the HPY Research Foundation through Elisa.}
\thanks{S.~Andreev and V.~Petrov are with Tampere University, Finland.}
\thanks{S.~Andreev and M.~Dohler are (also) with King's College London, UK.}
\thanks{H.~Yanikomeroglu is with Carleton University, Canada.}
}

\maketitle

\IEEEpeerreviewmaketitle

\begin{abstract}
For the past 40 years, cellular industry has been relying on static radio access deployments with gross over-provisioning. However, to meet the exponentially growing volumes of irregular data, the very notion of a cell will have to be rethought to allow them be (re-)configured on-demand and in automated manner. This work puts forward a vision of \textit{moving networks} to match dynamic user demand with network access supply in the beyond-5G cellular systems. The resulting adaptive and flexible network infrastructures will leverage intelligent capable devices (e.g., cars and drones) by employing appropriate user involvement schemes. This work is a recollection of our efforts in this space with the goal to contribute a comprehensive research agenda. Particular attention is paid to quantifying the network performance scaling and session continuity gains with ultra-dense moving cells. Our findings argue for non-incremental benefits of integrating moving access points on a par with conventional (static) cellular access infrastructure.
\end{abstract}

\section{Matching Access Supply and User Demand}

Today, mobile broadband systems have become a powerful operator asset to meet the rapid acceleration in the global traffic demand. The continued progress in user companion devices equipped with advanced computational intelligence and rich communication capabilities, such as smart phones, high-end wearables, connected vehicles, and autonomous drones, is however forcing network operators to respond promptly with decisive capacity scaling on their deployments~\cite{6736746}. The legacy fourth-generation (4G) wireless networks built around 3GPP LTE technology are already highly integrated and heterogeneous. However, the impending advent of user applications with stringent and highly-differentiated quality-of-service (QoS) requirements challenges the state-of-the-art 4G+ technologies with massive and variable volumes of traffic that nobody wants to pay for~\cite{7067426}. 

Therefore, it is commonly expected that the new fifth-generation (5G) networks will need to accommodate scenarios, which are not handled efficiently by the current cellular deployments. The global research on 5G radio access systems has essentially been concluded as 3GPP has ratified a new, non-backward-compatible ``New Radio'' (NR) technology in centimeter- and millimeter-wave (mmWave) spectra to augment further LTE evolution. However, we argue that even the novel 5G technology may face severe limitations due to unpredictable and non-uniform loading. Consequently, it may become insufficient to meet the QoS requirements of the end users, if supplied network capacity and demanded cell throughput do not match each other in space and time. 

Presently, the mainstream solution to mitigate the increasing disproportion between the irregular demand and the access supply is by deploying a higher density of heterogeneous small cells~\cite{6736747} in current cellular architecture. Notably, introducing an increasing number of serving stations is a gross over-provisioning also leading to more complex interference management and higher operator expenditures. In stark contrast, our work aims to conceptualize dynamic and flexible networks for ``on-demand'' densification.

\section{Ultra-Densification with 5G+ Moving Cells}

\begin{figure}[!ht]
  \centering
  \includegraphics[width=\columnwidth]{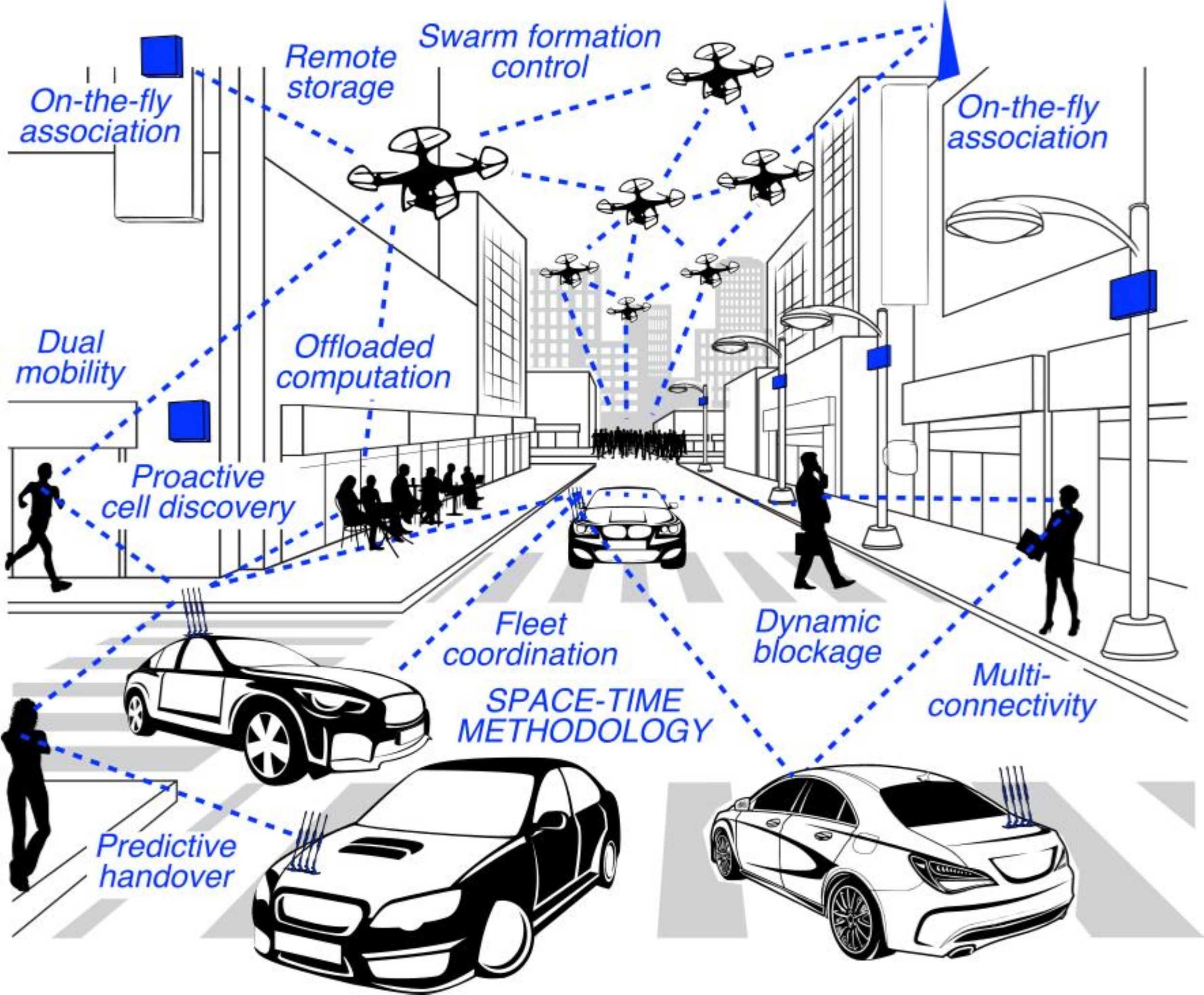}
  \caption{Utilization of moving cells in 5G+ networks.}
  \label{fig:concept}
\end{figure}

This section outlines our conceptual vision of a moving network in beyond-5G cellular infrastructure (see Fig.~\ref{fig:concept}).

Over the 40 years of its history, the cellular industry has primarily relied on static radio access network (RAN) deployments that may face serious limitations in mitigating the effects of dynamic, non-uniform loading. While access supply (i.e., the potential cell capacity) has been well-explored in the past~\cite{7010535}, the implications of user demand (i.e., the actual cell traffic) have only received very limited attention. In the near future, as more heterogeneous supply meets increasingly unpredictable demand, their mismatch in space and time threatens to create unprecedented levels of congestion. This is especially true for spontaneous large-scale events~\cite{7110500} that require service providers to augment the capacity of their networks quickly.

Increasingly widespread consumer devices, such as cars and drones, become a new breed of capable user equipment (UE) as they connect to the cellular broadband. While these user devices are rapidly expanding their functionality, the base stations are becoming smaller driven by the ongoing network densification. Hence, the original functional disparity between the network and the UE is becoming blurred, which offers new opportunities to utilize user devices as part of the network tasks. We thus expect the UE to take a more active role in 5G+ service provisioning and even occasionally assume the network infrastructure functions.

Intelligent UE may aid profoundly in providing wireless connectivity to relevant devices in proximity, such as offering device-to-device (D2D)-based data relaying, proximity gaming, content distribution, and caching. While D2D radio technology has originally been coined for public safety services~\cite{6163598}, it is likely to remain at the heart of the 5G ecosystem (named `sidelink'). In fact, engaging humans with their personal devices into collective activities is seen as the `last-resort' option for service operators, as it may bring a transformation from axiomatic network-centric to emerging device-centric system design. 

However, unpredictable and heterogeneous mobility of user devices may jeopardize session continuity and thus compromise the operator's service-level agreements. In urban environments, providing support for seamless connectivity becomes of paramount importance to improve the availability and reliability of wireless links for users and their connected devices moving at various speeds~\cite{6525596}. To achieve scalable and reliable content dissemination, the use of 5G-grade multi-connectivity in the context of truly mobile access has recently sparked renewed research interest. This calls for a careful investigation regarding the complex mixtures of mobility models and their effects. 

While multiple radio technologies and multi-access networks improve connection reliability, tolerance of link failures, and resource utilization, a question arises on the adequate sources of motivation that would facilitate the end-user decisions to lend their personal devices for collective usage~\cite{6616109}. In order to employ user-owned devices, operators need to embrace the selfish nature of humans: the decision strategy of the customers is inherently coupled with their utility perception. Hence, operator-driven mechanisms are strongly demanded to engage the masses of people into incentive-aware applications. 

Here, social interactions between an individual user and its proximal neighbors come into focus. With socially-aware cooperation, the emerging notion of ``social proximity'' empowers users to interact with the nearby matching people. Catering for tighter user involvement or directly impacting user behavior and decisions, the concept of ``user-in-the-loop''~\cite{6736762} enables people-aware applications. This disruptive paradigm extends the user's role beyond being a traffic-generating/-consuming `black box'. However, it also requires the development of unprecedentedly opportunistic and scalable radio network layouts.

\section{Moving Networks: Holistic Research Roadmap}

In this section, we contribute a concise summary of research challenges in moving access networks (MANs) for 5G+ wireless systems.

\subsection{Potential Performance of Dynamic Network Layouts}

Most past work in heterogeneous networks relies on stochastic geometry. However, as small cells are becoming even smaller, more heterogeneous, and mobile, this increased temporal dynamics requires an entirely new perspective. Here, queuing theory may be coupled with stochastic geometry to produce a hybrid methodology for characterizing space-time throughput regions at different levels of density. This converged approach can also be used for more efficient congestion control and optimized radio resource allocation across space and time. More complex but realistic cluster and hard-core spatial processes could be utilized here, which are more suitable to mimic the actual MAN deployments where nodes tend to group together and/or end up separated by some minimum distance. 

As new mmWave radio becomes a major 5G development, the particularities of extremely high frequency communication significantly complicate the above space-time formulations. When mmWave radio is used for MAN operation, the resulting blockage dynamics produced by various objects with dissimilar moving patterns (cars, drones, pedestrians) requires further careful research. The target here is to capture the intricate interplay between the feasible mmWave link length and blockage by human bodies or other mobile obstacles, as well as quantify the associated scaling laws. In this context, one could also employ random shapes theory that can model e.g., buildings with a random object process to complement past efforts based on random point processes.

Emergence of mmWave MAN infrastructures requires improved levels of radio link availability and reliability. While significant progress has been made in enhancing individual RATs, the ultimate potential of using heterogeneous multi-connectivity for higher communication reliability is not understood as of yet. New research is required to explore the gains made available with truly dynamic multi-radio integration, where individual radio links may remain unreliable, but adequate system-level performance reliability and service availability are guaranteed on top of them to satisfy the service-level agreements.

Recent studies confirm non-incremental benefits of proactive cell-edge data caching, both in user and network equipment, as well as call for quantifying the gains of remote computation offloading~\cite{7143326}. However, the majority of existing publications only consider these important developments individually, which limits the potential synergy. Future work will have to integrate communication, caching, and computing functionality (e.g., computing caches) for improved content availability. This includes methods for dynamic deployment of mobile computing caches. One should also consider the fundamental trade-off between content acquisition latency and node mobility to optimize placement of computing/caching nodes in future MAN architectures. 

\subsection{Practical Schemes for Moving Access Networks}

Today, most research studies are restricted to (semi-)static system topologies and/or may introduce prohibitive complexity for subsequent real-time implementation in MANs. The need for explicit treatment of movement in large fleets and formations (swarms) is being recognized only recently and the community needs to address the challenges of tackling heterogeneous mobility (produced by a mixture of dissimilar travel patterns), subject to appropriate cell moving models, such as e.g., Levy flight.

Utilization of future MANs in urban areas, especially over mmWave frequencies, requires detailed propagation and channel models in realistic city layouts. However, massive characterization of moving access points (MAPs) is cumbersome due to underlying complexity of capturing a large number of cells within accurate city maps. On top of this, spectrum aggregation and decoupled uplink/downlink mechanisms may be applied, but they have to be extended for higher moving speeds. The legacy core network architecture is a major bottleneck in cellular and currently its functionality is being pushed to the edge. One will need to explore distributed edge-network management and exploit decoupled/aggregated MAP connections.

Much prior literature assumes that reliable mmWave connectivity is only possible in the line-of-sight (LoS) conditions. This knowledge is important for constructing predictive handover solutions that are tailored for mmWave MAPs. Here, probabilistic LoS/nLoS information based on past channel statistics may be utilized proactively to improve session continuity and enhance end-to-end reliability. As a result, one has to construct smart handover solutions for dense mmWave systems with higher channel dynamics, where a user may need to perform more frequent cell transitions to compensate for short and highly directional mmWave links.

A complementary study is required on the minimum levels of control overhead with highly dynamic MAP changes at mmWave frequencies. This will enable the construction of efficient control algorithms, such as more reliable initial access protocols that employ signaling transmission redundancy to reduce beam acquisition delay. Viable tools for dealing with imperfect (non-ideal) control channels (e.g., capacity-limited, delayed, and with unreliable signaling) will need to be proposed. These should allow assessing the extent of minimal signaling overheads for efficient real-time MAP operation, as well as effectively balancing the developed intelligence between the user and the network equipment.

\subsection{Efficient Strategies to Integrate Moving Cells}

As MAPs integrate into the fabric of future networks, multiple coexistence-related questions will need to be resolved. First, one needs to address the aspects of how flying access points deployed on drones can affect ground networks. Indeed, while the moving patterns of cars can only be impacted to a limited extent (subject to appropriate driver incentives), the movement of drones may be steered based on the current supply--demand situation. However, the mobility of vehicles and humans is partially correlated, which may be exploited for opportunistic service provisioning. In this regard, a recent development to explicitly account for the end-user traffic activity and shape it opportunistically has been named user-in-the-loop (UIL). 

While network operators are forced to invest astonishing amounts into improving their network infrastructures, MANs target to actually impact user-generated traffic. In more detail, a user may receive suggestions and incentives (or even penalties) from the network when utilizing specific MAN applications/services, mindful of costs of using ground (e.g., cars) vs. flying (e.g., drones) MAPs. Hence, resource demand could be efficiently shaped in space and time as the association rules in MANs shift to one-on-one relationships due to extreme network densification. Ultimately, with the support from the cellular system, we expect that MAP operation can be automated, and devices may enjoy MAN benefits anytime/anywhere without considerable user overheads.

We further envision that engaging the increasingly capable user-owned equipment into collective utilization as part of MANs will soon become a viable option for the network operators to provide on-demand capacity, content, and coverage. This will however require appropriate pragmatic (e.g., monetary) user involvement mechanisms. One will need to develop the corresponding system-theoretic UIL models and apply them specifically to MANs, where the network will advise its clients on when and how to use the MAPs most effectively (e.g., based on current location, SINR, and traffic situation). This process can then be integrated into the client device, such that the user preferences are taken into account seamlessly.

We also expect that social structure of the network could be leveraged to reach the critical mass of user-provisioned MAN applications and services. However, to facilitate this vision, the very real security, privacy, and trust concerns need to be resolved. Here, the coexistence of closed vs. open access groups has to be addressed, especially in the cases of partial/no network coverage (e.g., edge of a cell, network failure, malicious attack), aiming to offer provable security and privacy mechanisms for MANs. Eventually, depending on the MAP mobility patterns, multiple applications and services may be suitable to run over MANs. We thus propose that the network is made aware of which content/service is requested by a client, using this knowledge throughout its operation.

\section{Practical Application of MAPs: A Case Study}

This section focuses on the system-level study of a MAN deployment.

\subsection{Characteristic Urban Setup and Deployment Parameters}

\subsubsection{Considered area of interest}

\begin{figure}[!ht]
  \centering
  \includegraphics[width=\columnwidth]{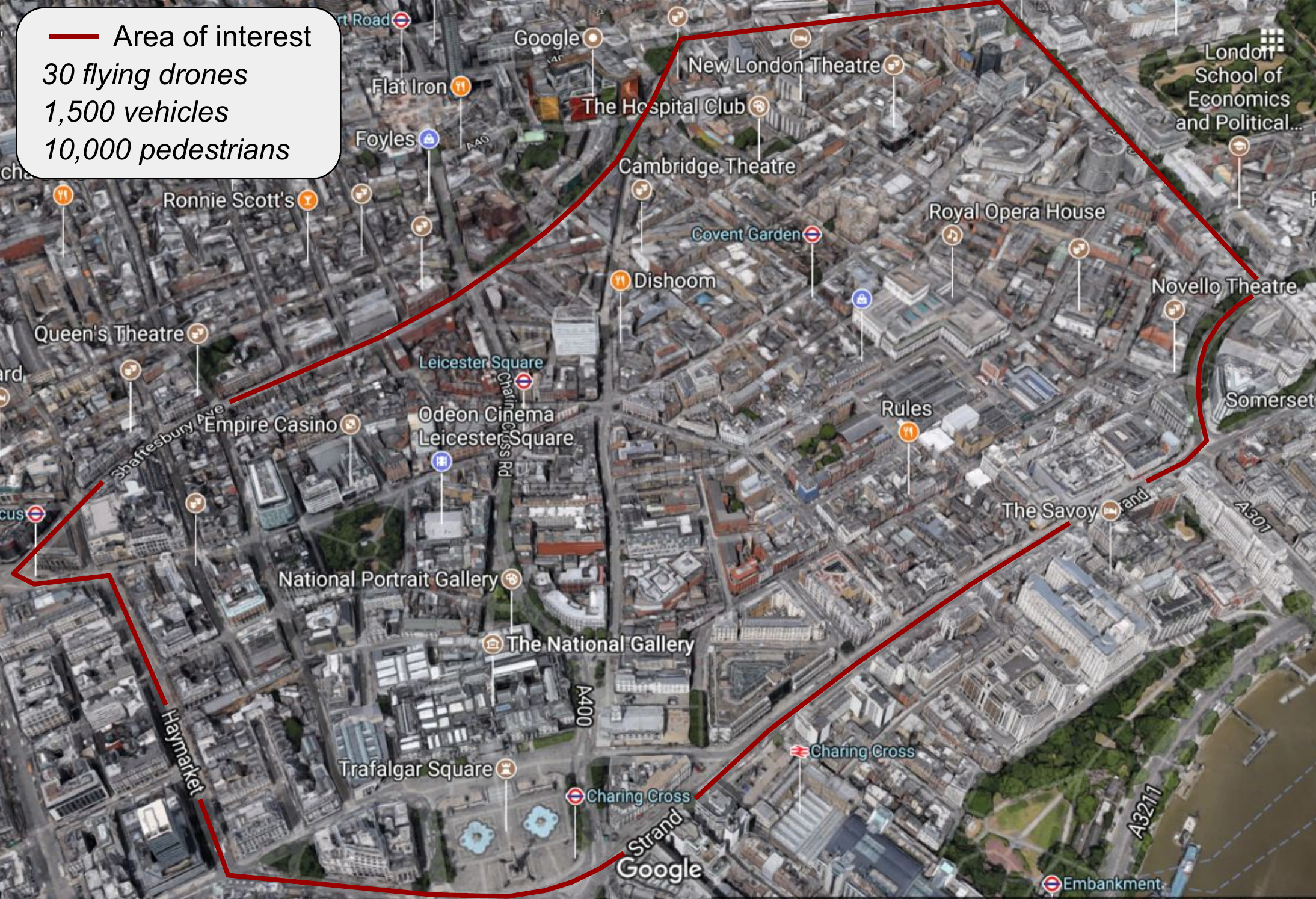}
  \caption{Our considered scenario of interest: part of central London, UK.}
  \label{fig:map}
\end{figure}

In this work, we consider a typical urban deployment. As a characteristic example, the area around King's College London, UK is adopted (see Fig.~\ref{fig:map}). This location is sufficiently representative for our purposes: it features wide avenues and small streets as well as includes large pedestrian zones (e.g., Trafalgar Square) and narrow sidewalks. Buildings are modeled as boxes of appropriate height, whereas smaller and temporary objects, such as lampposts and kiosks, are disregarded for simplicity.

\subsubsection{Pedestrians, vehicles, and drones}

Altogether, 10,000 pedestrians, 1,500 driving vehicles, and 30 flying drones are deployed in our scenario of interest simultaneously. The speed of drones, vehicles, and pedestrians is assumed constant and equals $40$\,km/h, $20$\,km/h (dense traffic), and $3$\,km/h, respectively. The vehicles as well as the pedestrians are assumed to follow the Manhattan mobility pattern: at every intersection, they select a new direction randomly, except for the direction that they were coming from. One of the available new directions is chosen with equal probability. Drones are assumed to travel across the area of interest with random and straight trajectories.

\subsubsection{Attractors and temporary crowds}

To capture the spatially and temporally correlated events, \textit{street performances} are modeled, which attract attention of the nearby pedestrians. A new event begins in a random location and at an arbitrary instant of time (on average, 10 events per hour). The duration of each street performance is set to $15$ minutes. A start of every such event makes $50\%$ of the pedestrians within view (i.e., inside $70$\,m range) alter their mobility patterns temporarily to form a crowd around the performers for the entire duration of a show. All of the spectators resume their mobility patterns once the performance is over.

\subsubsection{User traffic demand}

We assess a beyond-5G scenario where a notable fraction of humans actively employ \textit{augmented reality} (AR) glasses. In our setup, we assume the AR penetration to be at $10\%$, hence translating into nearly 1,000 connected AR gadgets in use. Upload of the multimedia capture from the event is considered as the main target application. We aim to specifically characterize the space-time surges in user traffic demand by assuming that each of the spectators wearing AR glasses decides to transmit a multimedia stream from the ongoing street performance.

\subsubsection{Deployed access network}

Given the massive offered load, 5G mmWave cellular is the only viable choice of an access network. Following the current 3GPP guidelines, mmWave access points are deployed on the walls of the buildings at $10$\,m height as a hexagonal grid with the inter-site distance of $200$\,m~\cite{3gpp_mmwave}. A share of vehicles/drones acting as MAPs -- named here the \textit{MAP involvement factor} -- varies from $0\%$ (no involvement, baseline scenario) to $100\%$ (full involvement, extreme case). Each of the vehicles/drones acting as a MAP is assumed to be equipped with the mmWave radio capabilities. Further deployment parameters and radio-specific settings are summarized in Table~\ref{tab:params}.

\begin{table}[!ht]
\caption{Main case-study parameters.}
\label{tab:params}
\begin{center}
\begin{tabular}{|p{0.25\columnwidth}|p{0.65\columnwidth}|}
\hline
\multicolumn{2}{|l|}{\textbf{Deployment}}\\
\hline
Area of interest & \textbf{Location:} London, UK\\
& \textbf{Area:} Trafalgar Square\\
& \textbf{Size:} $\approx0.5$\,km$^2$\\
& \textbf{Buildings:} From real map with actual heights\\
\hline
Street events & \textbf{Starting location:} Uniform within the area\\
& \textbf{Attracted pedestrians:} $50\%$ within $70$\,m\\
& \textbf{Periodicity:} $\approx10$ per hour\\
& \textbf{Duration:} $15$\,min (a single performance) \\
\hline
Pedestrians & \textbf{Number:} 10,000 per area\\
& \textbf{Model:} Cylinder $1.7$\,m height, $0.5$\,m width\\
& \textbf{Speed:} $3$\,km/h (constant)\\
& \textbf{Mobility:} Manhattan pattern on the sidewalks\\
\hline
Vehicles & \textbf{Number:} 1,500 per area\\
& \textbf{Model:} Parallelepiped $4.8$\,m $\times$ $1.8$\,m $\times$ $1.4$\,m\\
& \textbf{Speed:} $20$\,km/h (constant) \\
& \textbf{Mobility:} Manhattan pattern on the roads\\
\hline
Drones & \textbf{Number:} 30 per area\\
& \textbf{Height:} $60$\,m\\
& \textbf{Speed:} $40$\,km/h (constant)\\
& \textbf{Mobility:} Random straight flight across the area\\
\hline
\multicolumn{2}{|l|}{\textbf{User gadgets}}\\
\hline
Devices & \textbf{Category:} AR glasses\\
& \textbf{Number:} About 2,000 in the area ($10\%$ of people)\\
\hline
User traffic & \textbf{Service:} Uplink multimedia streaming\\
& \textbf{Video quality:} 2K with $30$\,fps\\
& \textbf{Stream duration:} Exponential with the mean of $5$\, min\\
& \textbf{Rate of streams:} $\approx3$ streams per hour\\
& \textit{Users capture street performance in their proximity}\\
\hline
\multicolumn{2}{|l|}{\textbf{Radio access network}}\\
\hline
mmWave radio & \textbf{Frequency:} $28$\,GHz\\
& \textbf{Bandwidth:} $1$\,GHz\\
& \textbf{Propagation:} 3GPP UMi -- Street canyon~\cite{3gpp_mmwave}\\
& \textbf{Effect of buildings:} LoS $\rightarrow$ nLoS\\
& \textbf{Effect of humans/vehicles:} nBlocked $\rightarrow$ Blocked\\
& \textbf{Signal degradation with blockage:} $20$\,dB~\cite{blockage_20db_paper}\\
\hline
Static mmWave & \textbf{Deployment:} Hexagonally arranged cell sites\\
access points& \textbf{Density:} from $20$\,APs/km$^2$ to $250$\,APs/km$^2$\\
& \textbf{Sectorization:} 3 sectors per site\\
& \textbf{Downtilting:} $102\degree$ from the vertical axis\\
& \textbf{Transmit power:} $35$\,dBm\\
& \textbf{Height:} $10$\,m\\
\hline
Vehicular mmWave & \textbf{Transmit power:} $30$\,dBm\\
MAPs (vMAPs) & \textbf{Height:} $1.4$\,m (vehicle's roof)\\
\hline
Aerial mmWave & \textbf{Transmit power:} $23$\,dBm\\
MAPs (aMAPs) & \textbf{Height:} $60$\,m (drone's bottom)\\
\hline
User equipment & \textbf{Transmit power:} $20$\,dBm\\
(AR glasses)&  \textbf{Height:} $1.65$\,m (eye level)\\
\hline
\end{tabular}
\end{center}
\end{table}

\subsection{Summary of Conducted System-Level Evaluations}

\subsubsection{Developed simulation framework}

To characterize the described scenario and produce first-order performance evaluation, an in-house system-level simulation framework was employed. It is implemented in Python and operates in a time-driven fashion. More specifically, the target area of interest is processed to construct two path graphs: one for MAPs and another one for pedestrian users. For the sake of better accuracy in the output results, all of the collected intermediate data are averaged over $20$ replications. Each of such replications corresponds to one hour of real-time operation.

\subsubsection{Modeling mmWave radio}

We focus on the currently ratified 5G mmWave cellular technology operating at the carrier frequency of $28$\,GHz with $1$\,GHz of bandwidth. The radio links between all of the users are modeled based on the \emph{UMi -- Street canyon} path loss model~\cite{3gpp_mmwave}, where LoS and nLoS conditions are differentiated subject to the relative positions of nodes as well as the obstacles between them (such as buildings). The LoS conditions may be further susceptible to random blockage that occurs whenever a link between the communicating nodes is occluded by a vehicle or a human. 

\subsubsection{User association, handover, and backhaul}

Any mmWave access point (including car- and drone-mounted MAPs) accepts a new user session only if it has sufficient radio resources to handle it, while the MAPs also ensure that their backhaul capacity is sufficient. Access and backhaul connections of MAPs coexist in the spatial and frequency domains, such that the radio resources of an access point are shared dynamically between them. Below, we differentiate between user- and network-controlled user association policies:

\begin{itemize}
\item \textbf{User-controlled association.} Whenever a new session arrives, the initiating user selects an access point with the highest signal power conditioning on the fact that it has adequate radio resources to handle this session. In the course of its multimedia streaming, the user continuously monitors other potential access points in proximity. Should one with better signal strength and sufficient resources become available, the ongoing user session transfers to it seamlessly by leveraging multi-connectivity capabilities~\cite{7556954}.

\item \textbf{Network-controlled association.} Whenever a new session arrives, the network analyzes all of the feasible beam configurations in the area, such that a certain target metric of interest is optimized. The necessary changes in the corresponding connectivity graph are then enforced for both the user with its new session and any other user, which needs to change its serving access point by following the network instructions. The handover between access points is performed similarly to that in user-controlled association.
\end{itemize}

From the above, the former policy represents an example of the locally-optimal (`greedy') solution, while the latter reflects achievable bounds for the outage probability, since the network is assumed to have better knowledge of all its parameters, possess superior computation capabilities, and enjoy optimized signaling procedures.

\setcounter{subsubsection}{0}
\section{Combating Outage: MAN Performance Scaling}

\subsubsection{Parameters and target metrics of interest}
	
Past studies of small cell networks focus primarily on capacity scaling and interference characterization~\cite{doi:10.1002/ett.3145}. In this work, we argue for the importance of \emph{individual capacity share} and \emph{outage probability} as two representative performance indicators in urban environments, which are also crucial for the system operators in their network planning. Here, we define the outage probability as the chances that the target user experiences outage with respect to its selected access point (SINR $<$ 3\,dB) when a session starts. The term individual capacity share reflects the proportion of the total cell capacity in bits/s that an average user receives in the considered scenario. In our study, we model joint access and in-band backhaul operation.

\subsubsection{Outage probability gain with MAPs}

\begin{figure}[!ht]
  \centering
  \includegraphics[width=\columnwidth]{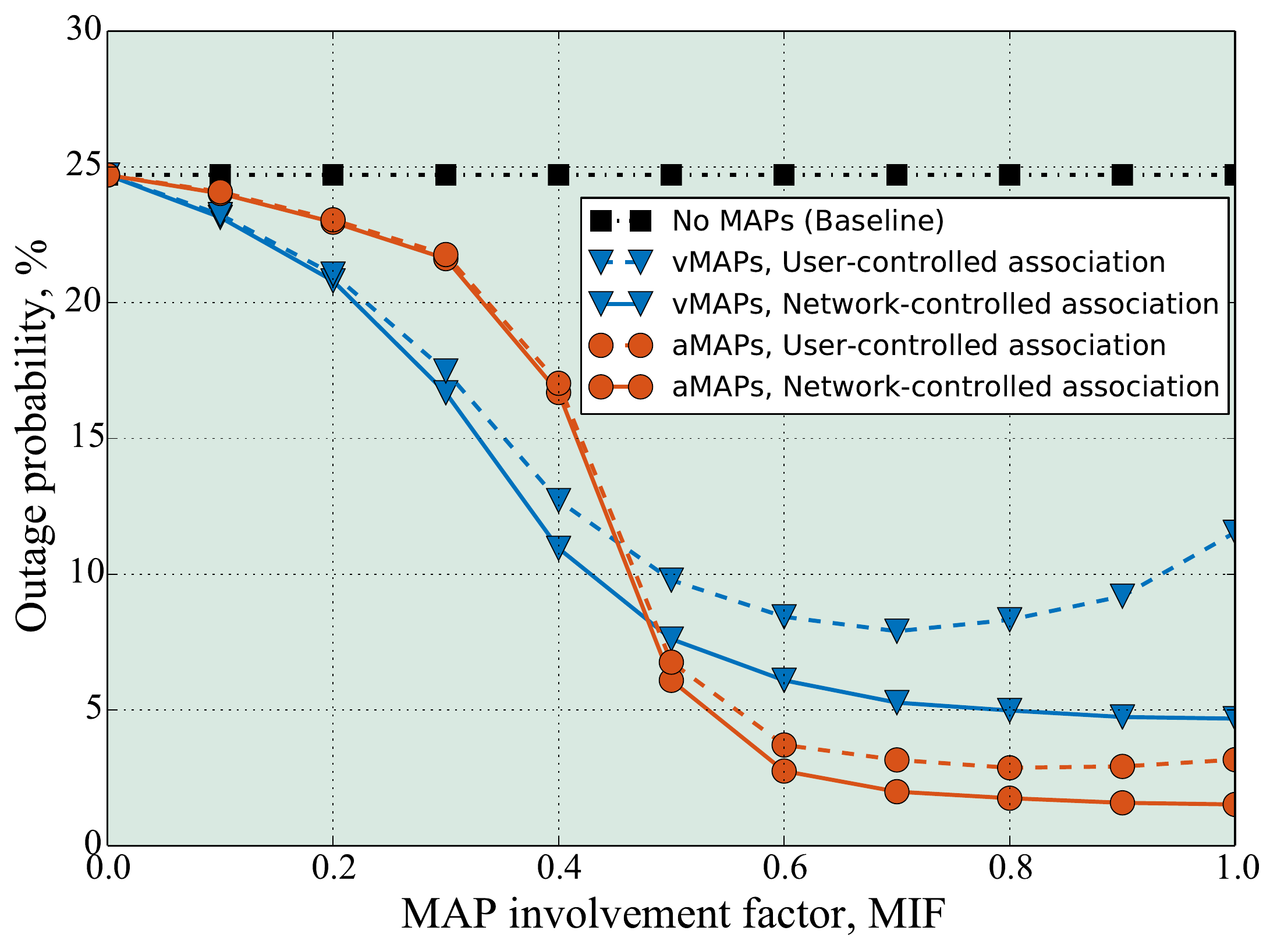}
  \caption{User benefits. Outage probability gain with MAP involvement.}
  \label{fig:plot1}
\end{figure}

First, Fig.~\ref{fig:plot1} reports on the outage probability as a function of the MAP involvement factor (MIF). The density of static mmWave APs is set to $40$ per km$^2$. Clearly, the utilization of MAPs rapidly improves the outage performance and helps significantly outperform the baseline system operation especially for higher MIF values. The following important effects are thus observed.

\textbf{Effects of MAP involvement factor.} At lower involvement factors (MIF $<$ 0.4), vMAPs are preferable since there are but a few drones acting as aMAPs. Hence, involving a relatively small proportion of driving vehicles (e.g., taxi fleet cars) with MAP capabilities readily leads to notable benefits in user service. At higher involvement levels (MIF $>$ 0.5), the use of aMAPs is preferred as these are less susceptible to mmWave blockage and interference. In other words, utilization of e.g., third-party drones as MAPs is sensible only starting from their notable numbers (i.e., tens per sq. km).

\textbf{Effects of MAP association strategy.} User-controlled association with vMAPs operates reasonably well for lower involvement factors (MIF $<$ 0.3). At higher involvement levels (MIF $>$ 0.3), more intelligent network-controlled association solutions are preferred since user-controlled association faces increasingly stringent interference and backhaul constraints. Interestingly, the performance of vMAPs with user-controlled association degrades slightly for the very high MIF values. However, this effect is not critical for aMAPs (only notable for MIF $>$ 0.8) due to much lower interference in this case. Hence, simpler user-controlled association is suitable for drone-based mmWave access.

\subsubsection{Trading static mmWave access for MAPs}

\begin{figure}[!ht]
  \centering
  \includegraphics[width=\columnwidth]{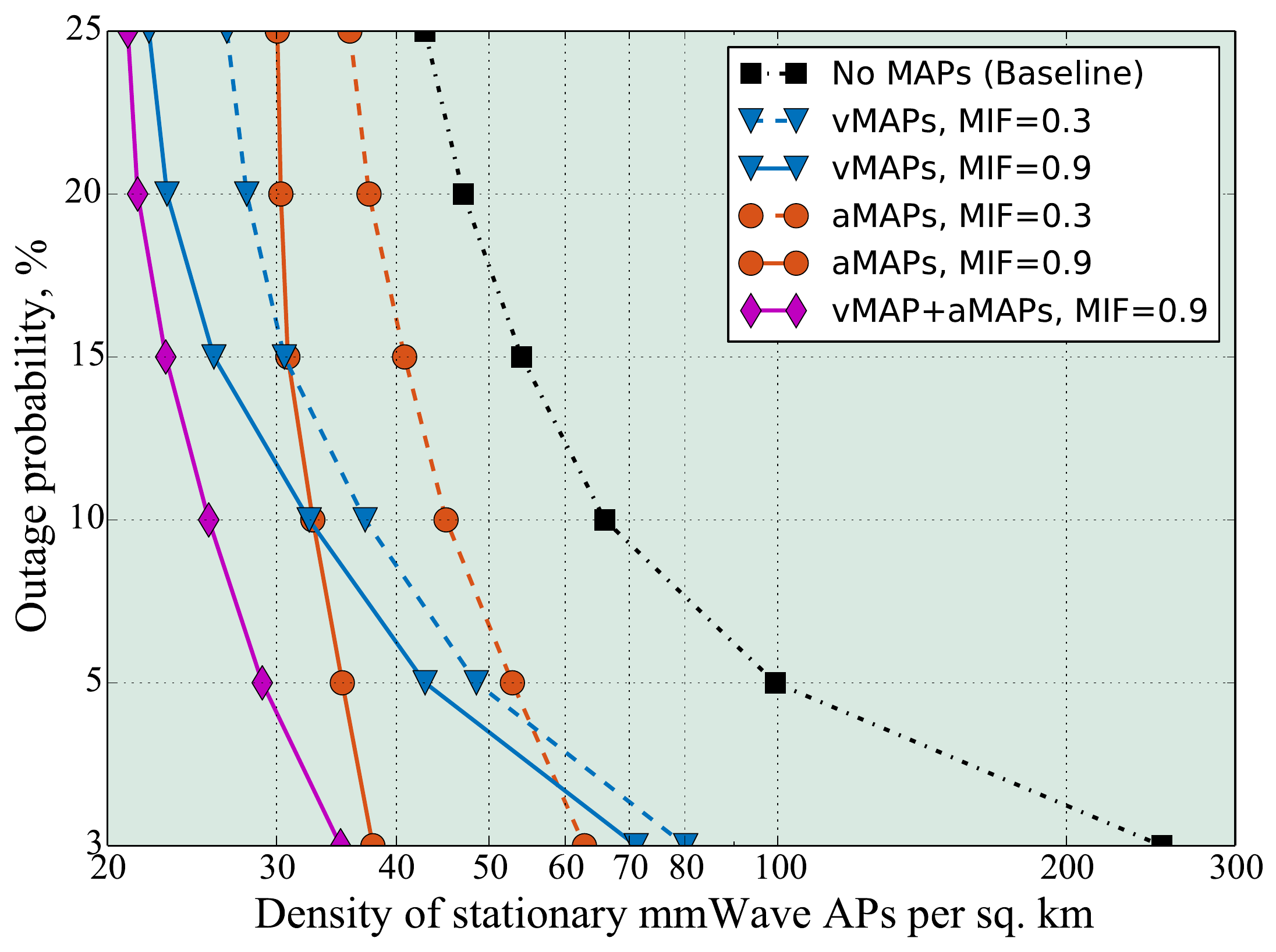}
  \caption{Operator benefits. Savings on static mmWave infrastructure.}
  \label{fig:plot2}
\end{figure}

\begin{figure}[!ht]
  \centering
  \includegraphics[width=\columnwidth]{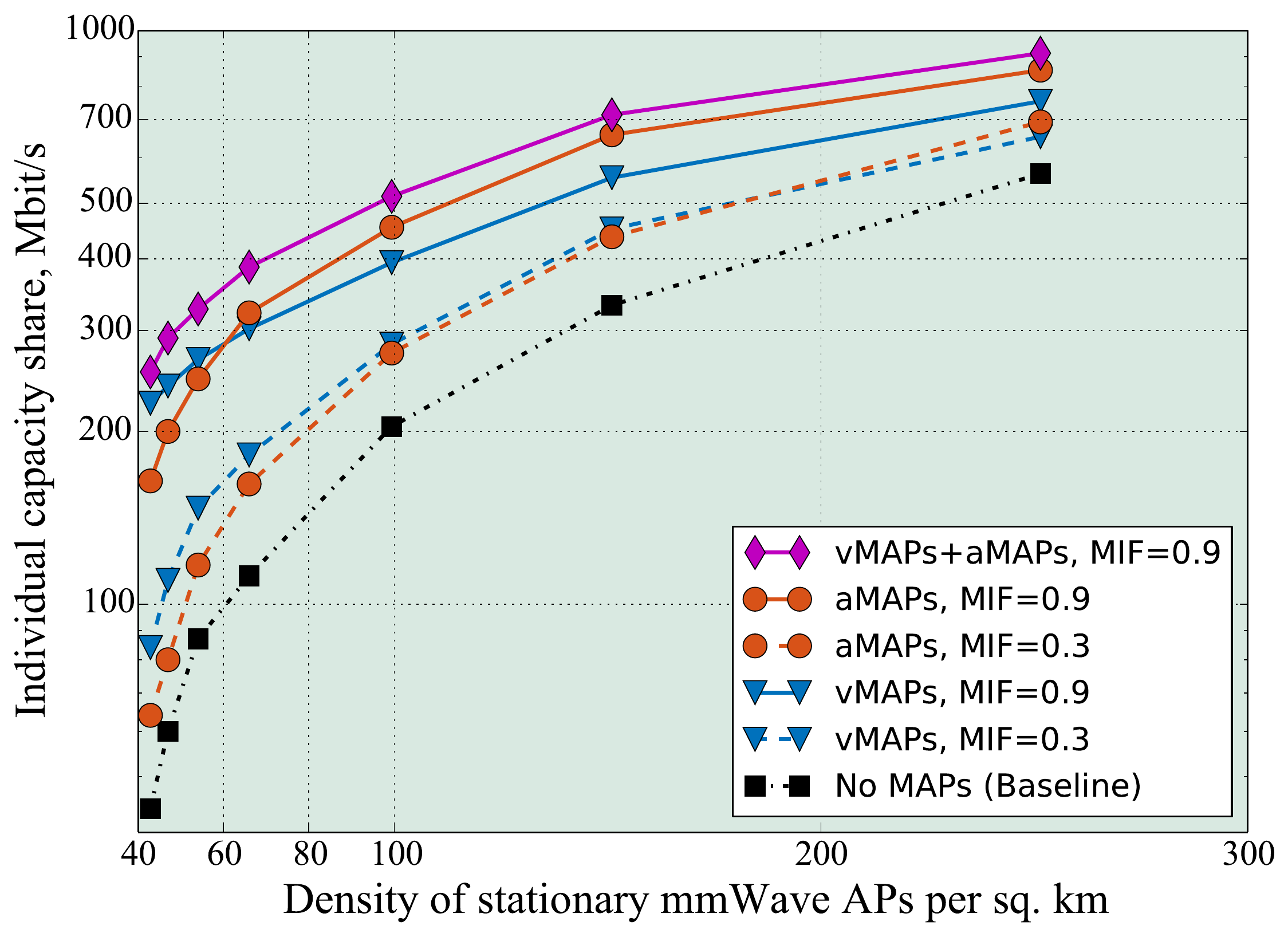}
  \caption{User benefits. Capacity gain with MAP involvement.}
  \label{fig:plot3}
\end{figure}

Further, deploying MAPs allows to significantly decrease the required density of static mmWave access nodes to maintain certain outage and capacity levels. This is confirmed by Fig.~\ref{fig:plot2} and Fig.~\ref{fig:plot3}, where the network-controlled association is considered and the following effects are observed.

\textbf{Effects of MAP penetration.} In Fig.~\ref{fig:plot2}, the impact of MIF is more pronounced for vMAPs at higher outage levels (difference between MIF = 0.3 and MIF = 0.9 is high). However, lower outage probabilities (e.g., $Pr\{Outage\}$ = $3$$\%$) are difficult to achieve even with sufficiently high penetration of vMAPs due to poor channel conditions around the street events. A somewhat opposite situation is observed for aMAPs, where the strongest impact of MIF is seen when serving outage-constrained applications.

\textbf{Effects of capacity with MAPs.} In its turn, Fig.~\ref{fig:plot3} reports that for the lower densities of static mmWave APs, the vMAPs are preferable since aMAPs often do not have an available static AP in proximity to reliably backhaul their traffic to. Meanwhile, at its certain density, a backhaul-limited aMAP network transitions to access-limited regime (see Fig.~\ref{fig:plot3}) and begins outperforming the use case with vMAPs.

\textbf{Effects of static AP density.} While similar performance in terms of the outage probability and individual capacity share can be achieved without MAPs (see the \emph{Baseline} operation in Fig.~\ref{fig:plot2} and Fig.~\ref{fig:plot3}), this requires a significantly higher density of static mmWave APs, thus escalating the capital and operational expenditures. Our results suggest that -- if the MAP deployment and maintenance costs permit -- certain ultra-dense deployments of static mmWave APs can be replaced by a combination of static and MAP-based layouts, by bringing further flexibility and potential savings to service providers.

\subsubsection{Aerial and vehicular heterogeneous MAP} 
Finally, the impact of utilizing both vMAPs and aMAPs is discussed (see the magenta curves in Fig.~\ref{fig:plot2} and Fig.~\ref{fig:plot3}).

\textbf{Effects of target outage.} In Fig.~\ref{fig:plot2}, vMAPs are preferred for higher target outage levels ($Pr\{Outage\}$$>$$10$$\%$) since even their low penetration (MIF = 0.3) results in hundreds of assisting car-cells. Meanwhile, lower target outage values ($Pr\{Outage\}$$\approx$$3$-$5$$\%$) can hardly be achieved without the utilization of aMAPs as only these can reliably connect e.g., users in the center of a crowd that surrounds the street event. 

\textbf{Effects of heterogeneous MAPs.} In conclusion, the joint utilization of vMAP and aMAP becomes more beneficial since vehicles and drones offer complementary benefits~\cite{8473484}. Given that they are less vulnerable to human-body blockage, the key role of aMAPs in this setup is to serve dense user crowds, whereas the numbers of vMAPs are sufficient to accommodate the remaining consumers distributed across the area of interest. The aggregate gains of this \textit{heterogeneous} MAN system range from 20$\%$ to 90$\%$ depending on the assumed density of static mmWave APs.

\section{Utilization of MAPs in Beyond-5G Ecosystem}

Mass adoption of multi-gigabit mmWave access infrastructures is expected to decisively augment network capabilities in handling accelerated traffic with stringent throughput and reliability demands. The forthcoming 5G systems will therefore have sufficient capacity to accommodate the \emph{median loading}. However, even the advanced 5G mmWave cellular technology may become insufficient to cope with spatially-correlated traffic bursts produced by unexpected mass events, such as spontaneous outdoor gatherings. In these situations, the emerging augmented and virtual reality applications may challenge the network with multiple high-resolution multimedia streams. This pushes network operators to further densify their deployments, which may be done quickly with the discussed concept of moving cells.

A prominent future research direction is highlighting the advantages of MAN architectures in light of the latest industrial efforts on small cell topology with fixed backhaul wireless, such as Facebook Terragraph, Facebook Aquila, Google Loon, 3GPP and IEEE efforts on small cells, etc. Our envisaged moving access infrastructures will offer mobile operators an opportunity to dramatically boost system capacity in their desired area of interest \emph{on demand}, by dynamically engaging a sufficient number of proximate MAPs. The performance results contributed in this work evidence that MAPs deliver the much needed capacity and session continuity improvements, which surpass the performance of conventional (static) mmWave cellular access.

\bibliographystyle{ieeetr}
\bibliography{moving_infrastructure}

\section*{Authors' Biographies}

\textbf{Sergey Andreev} (sergey.andreev@tuni.fi) is an assistant professor of electrical engineering at Tampere University, Finland. Since 2018, he has also been a Visiting Senior Research Fellow with the Centre for Telecommunications Research, King's College London, UK. He received his Ph.D. (2012) from TUT as well as his Specialist (2006) and Cand.Sc. (2009) degrees from SUAI. He (co-)authored more than 150 published research works on intelligent IoT, mobile communications, and heterogeneous networking.

\textbf{Vitaly Petrov} (vitaly.petrov@tuni.fi) is a Ph.D. candidate at Tampere University, Finland. He received the Specialist degree (2011) from SUAI University, St. Petersburg, Russia and the M.Sc. degree (2014) from Tampere University of Technology, Finland. He was a Visiting Scholar with Georgia Institute of Technology, Atlanta, USA, in 2014. Vitaly (co-)authored more than 30 published research works on millimeter-wave/terahertz band communications, Internet-of-Things, nanonetworks, cryptology, and network security.

\textbf{Mischa Dohler} (mischa.dohler@kcl.ac.uk) is a full professor in Wireless Communications at King's College London, driving cross-disciplinary research and innovation in technology, sciences, and arts. He is a Fellow of the IEEE, the Royal Academy of Engineering, the Royal Society of Arts (RSA), the Institution of Engineering and Technology (IET); and a Distinguished Member of Harvard Square Leaders Excellence. He is a serial entrepreneur; composer \& pianist with 5 albums on Spotify/iTunes; and fluent in 6 languages.

\textbf{Halim Yanikomeroglu} (halim@sce.carleton.ca) is a full professor in the Department of Systems and Computer Engineering at Carleton University. His research interests cover many aspects of wireless technologies with special emphasis on cellular networks. His collaborative research with industry has resulted in 27 granted patents. He is a Fellow of the IEEE and the Engineering Institute of Canada (EIC), a Distinguished Speaker for both the IEEE Communications Society and the IEEE Vehicular Technology Society.

\end{document}